\documentstyle[12pt]{article}
\def\vins{\mathbin{\dimen0=\ht\strutbox  \divide\dimen0 by 2
\hbox{\vbox{\hrule width\dimen0}\hskip-0.4pt\vrule height\dimen0}}\,}

\def\tIH{\tilde{\IH}}

\def\tS{\tilde{\tS}}
\def\re{{\bf R}}

\def\const{{\rm const}}
\def\vol{\,{}^2\!\epsilon}

\def\be{\begin{equation}}
\def\ee{\end{equation}}
\def\ba{\begin{eqnarray}}
\def\ea{\end{eqnarray}}
\def\L{{\cal L}}
\def\D{{\cal D}}
\def\Lie{\L}

\def\Re{{\rm Re}}
\def\Im{{\rm Im}}

\def\a{{\alpha}}

\def\b{{\beta}}

\def\kl{\kappa^{(\ell)}}

\def\m{m^a\partial_a}
\def\mb{\bar{m}^a\partial_a}

\def\pullback{\hat{=}}
\def\={\pullback}

\def\IH{\Delta}
\def\hIH{\hat{\IH}}

\def\tIH{\tilde{\IH}}

\def\tS{\tilde{\tS}}
\def\re{{\bf R}}
\def\IH{{\cal \Delta}}

\def\const{{\rm const}}
\def\be{\begin{equation}}
\def\ee{\end{equation}}
\def\ba{\begin{eqnarray}}
\def\ea{\end{eqnarray}}
\def\L{{\cal L}}
\def\Lie{\L}

\def\Re{{\rm Re}}
\def\Im{{\rm Im}}

\def\a{{\alpha}}

\def\b{{\beta}}

\def\kl{\kappa^{(\ell)}}

\def\m{\delta}
\def\mb{\bar{\delta}}

\def\pullback{\hat{=}}
\def\={\pullback}

\def\IH{\Delta}

\def\tIH{\tilde{\IH}}

\def\tS{\tilde{\tS}}
\def\re{{\bf R}}
\def\IH{{\cal \Delta}}

\def\const{{\rm const}}
\def\be{\begin{equation}}
\def\ee{\end{equation}}
\def\ba{\begin{eqnarray}}
\def\ea{\end{eqnarray}}
\def\L{{\cal L}}
\def\Lie{\L}

\def\Re{{\rm Re}}
\def\Im{{\rm Im}}

\def\a{{\alpha}}

\def\b{{\beta}}

\def\kl{\kappa^{(\ell)}}

\def\m{\delta}
\def\mb{\bar{\delta}}

\def\pullback{\hat{=}}
\def\={\pullback}
\def\IH{\Delta}

\def\q{\hat{q}}
\def\pback#1{{
\mathchoice{\StemPullBack{#1}{\leftarrowfill}}
     {\StemPullBack{#1}{\leftarrowfill}}
             {\IndxPullBack{#1}{\leftarrowfill}}
         {\IndxPullBack{#1}{\leftarrowfill}}}\vphantom{#1}}

\newcommand{\StemPullBack}[2]{
  \vtop{\mathsurround=0pt
  \ialign{##\crcr$\textstyle{#1}\strut$\crcr
    \noalign{\kern-0.4ex\nointerlineskip}{\tiny#2}\crcr}}}

\newcommand{\IndxPullBack}[2]{
  \vtop{\mathsurround=0pt
  \ialign{##\crcr\hfil$\scriptstyle{#1}$\hfil\crcr
    \noalign{\kern+0.4ex\nointerlineskip}{\tiny#2}\crcr}}}
\begin{document}
\title{Geometric Characterizations of the Kerr Isolated Horizon.}
\author{Jerzy Lewandowski${}^{1,2}$ and Tomasz Pawlowski${}^{1,2}$}
\date{}
\maketitle
\centerline{\it 1. Instytut Fizyki Teoretycznej, Wydzial Fizyki,
Uniwersytet Warszawski,} \centerline{\it ul. Hoza 69,
00-681,Warszawa, Poland}
\centerline{\it 2. Max Planck Institut f\"ur Gravitationsphysik,}
\centerline{\it Am Muehlenberg 1, D-14476 Golm, Germany}
\begin{abstract}
We formulate  conditions on the geometry of a non-expanding
horizon $\Delta$ which are sufficient for the space-time metric to
coincide on $\Delta$ with the Kerr metric. We introduce an
invariant which can be used as a measure of how different the
geometry of a given non-expanding horizon is from the geometry of
the Kerr horizon. Directly, our results concern the space-time
metric at $\IH$ at the zeroth and the first orders. Combained with
the results of Ashtekar, Beetle and Lewandowski, our conditions
can be used to compare the space-time geometry at the
non-expanding horizon with that of Kerr to every order. The
results should be useful to numerical relativity in analyzing the
sense in which the final black hole horizon produced by a collapse
or a merger approaches the Kerr horizon.
\end{abstract}

\section{\bf Introduction.}
In a new quasi-local theory of black holes
\cite{AFK,ABDFKLW,ABL1,ABL2}, one considers situation in which the
black hole has reached equilibrium although the exterior
space-time still admits outgoing radiation. The black hole in
equilibrium is described by a non-expanding horizon, i.e. a null
cylinder $\IH$, generated by segments of null geodesics orthogonal
to a space-like 2-surface diffeomorphic with a 2-sphere.
The geometry of $\IH$ relevant for extracting physics of $\IH$ is
defined by the space-time metric tensor $g_{ab}{}_{\mid_\IH}$ and its
derivative $\Lie_n g_{ab}{}_{\mid_\IH}$ with respect to a transversal
(also called radial) vector field $n$. (In the next section we recall
the definition which does not depend on the choice of an $n$.) The
geometry of $\IH$ has local degrees of freedom even if we assume that
the vacuum Einstein equations hold in a neighborhood of $\IH$. A
priori the Kerr solution does not appear to play a special role in
this context: the Kerr horizon is only an example of a non-expanding
horizon. On the other hand, there are some heuristic arguments
suggesting that a black hole formed in a physical process should
converge, in some suitable sense, to the Kerr black hole \cite{A}. To
probe this important issue of the `final state', one can begin with a
preliminary question: what is condition on the geometry of a
non-expanding horizon $\IH$ that ensures that its geometry coincides
with that of the Kerr horizon? We will analyze this question
here. 

This analysis will not unravel any unknown properties of the Kerr
metric. Rather, our goal is to select a covariantly defined property
of the Kerr horizon which uniquely distinguishes its geometry among
all non-expanding horizons.  Our conditions are local to $\IH$ in
contrast to characterizations of the Kerr solution that rely on global
assumptions about space-time and can be easily checked in numerical
simulations. The detailed calculations and proofs will appear in
\cite{LP}. 

In Section \ref{A} we explain how our local
characterization can be used to compare the space-time metric tensor
on a given non-expanding horizon with the Kerr solution to every order
in an expansion with respect to a coordinate parametrizing incoming
null geodesics transversal to $\IH$.

\section{Isolated horizons.}
\centerline{\it Definitions}

A {\it non-expanding horizon} is a null surface $\IH$ in a
4-spacetime $M$ such that

\begin{itemize}
\item{} $\IH$ is generated by  segments of null geodesics orthogonal to
a space-like 2-sphere $\tIH\subset M$, and is diffeomorphic to
$\tIH\times \re$;

\item{} the expansion of every null vector field $\ell^a$ tangent to $\IH$
vanishes;

\item{} Einstein's equations hold on $\IH$ and the stress-energy
tensor $T_{ab}$ is such that $-T^a{}_{b}\ell^b$ is future pointing
for every null vector tangent to $\IH$.

\end{itemize}

It follows from the conditions above, that the degenerate metric
tensor $q_{ab}=g_\pback{ab}$ is Lie dragged by $\ell^a$,
\be\label{1} \Lie_\ell q_{ab}\ =\ 0. \ee

The parallel transport defined by the space-time connection along
any curve contained in $\IH$ preserves the tangent bundle
$T(\IH)$, and induces a connection $\D_a$ therein. The pair
$(q_{ab}, \D_a)$ is referred to as  {\it the geometry of} $\IH$.
If a non-expanding horizon $\IH$ admits a null vector field
$\ell^a$ such that its flow $[\ell]$ is a symmetry of
$(q_{ab},\D_a)$, then we say that $(\IH, [\ell])$ is an {\it
isolated horizon}. We also assume that the restriction of the flow
$[\ell]$ to every null geodesic is non-trivial.

An isolated horizon $(\IH, [\ell], q_{ab}, \D_a)$ defined by the
non-extremal future, outer (inner) event horizon of the Kerr
metric, with $\ell^a$ the restriction to $\IH$ of the Killing
vector field which is null at $\IH$, will be called {\it the Kerr
outer  (inner) isolated horizon}.

\medskip

\centerline{\it The degrees of freedom.}

The degrees of freedom in the geometry of an isolated horizon are:
the intrinsic metric tensor $q_{ab}$, the {\it rotation 1-form
potential} $\omega_a$ defined on $\IH$ by the derivative of
$\ell^a$, \be \D_b\ell^a\ =\ \omega_b\ell^a, \ee and the pull-back
$R_{\pback{ab}}$ of the Ricci tensor constrained by \cite{AFK} \be
R_{\pback{ab}}\ell^b=0. \ee Owing to the definition of an isolated
horizon, we have \be\label{sym}  \L_\ell \omega_a\ =\ \L_\ell
R_{\pback{ab}}\ =\ 0. \ee The factor $\kl$ in $\ell^a\D_a\ell^b\
=\ \kl\ell^b$ is called {\it the surface gravity} of $\ell^b$. It
follows from the symmetry of the isolated horizon and from the
vanishing of $R_{\pback{ab}}\ell^b$, that the {\it surface
gravity} $\kl$ is constant, \be \kl\ =\ \const. \ee We call an
isolated horizon {\it non-extremal} whenever \be \kl\neq 0, \ee
and {\it extremal} otherwise.
%

An isolated horizon $\IH$ will be called {\it vacuum isolated
horizon} if the pull-back $R_{\pback{ab}}$ of the Ricci tensor
vanishes on $\IH$. It will be useful to characterize the geometry
of a non-extremal vacuum isolated horizon $(\IH, [\ell])$  by two
scalar invariants \cite{ABL2}. The first one is the Gauss
curvature $K$ of the 2-metric $\tilde{q}$ induced on any
space-like 2-surface passing through a given point of $\IH$. The
second one is the rotation scalar $\Omega$ defined by \be d\omega\
=\ \Omega\, \vol, \ee where $\vol$ is the area
2-form\footnote{There is naturally defined area 2-form $\vol$ on
each null 3-surface such that for every space-like 2-subsurface
$S'$ the integral $\int_{S'}\vol$ equals the area of $S'$.} of
$\IH$. The invariants are combined into a single, complex valued
function, namely the component $\Psi_2 = C_{abcd}\ell^a n^b(\ell^c
n^d - m^c \bar{m}^d)$ of the Weyl tensor, where $m^a,n^a$ are any
complex and, respectively, real null vector fields such that
$n_a\ell^a=-1$, $m_a\bar{m}^a=1$ and $m_an^a=m_a\ell^a=0$; we have
\be \Psi_2\ =\ \frac{1}{2}\big(-K + i\Omega\big). \ee For every
global, space-like cross-section $\tilde{\IH}$ of $\IH$, $\Psi_2$
satisfies the following global constraint, \be \int_{\tIH} \Psi_2
\vol\ =\ -2\pi.\label{int} \ee Every non-extremal isolated horizon
geometry $(\IH, q_{ab},\D_a,[\ell])$ is
determined\footnote{Isolated horizon is not assumed to be
geodesically complete. Therefore, some extra information would be
needed to know which finite segment of each null geodesics tangent
to $\IH$ is contained in $\IH$.}, up to diffeomorphisms preserving
the null generators of $\IH$ by the pair of invariants $(K,\,
\Omega)$. Owing to the symmetry $(\ref{1},\,\ref{sym})$, the
invariant $\Psi_2$ is constant along the null generators of $\IH$.
Therefore, it defines a function on the sphere $\hIH$ of the null
geodesics of $\IH$. It follows from the results of
\cite{Lewandowski}, that given a 2-sphere $\hIH$ equipped with a
metric tensor $\q_{AB}$ and a function $\hat{\Omega}$ such that
$\int_{\hIH} \hat{\Omega}\, {}^2\!\hat{\epsilon}\ =0$, there is a
non-extremal vacuum isolated horizon whose invariants $K$ and
$\Omega$ correspond to $\q$ and $\hat{\Omega}$.

\section{Geometric conditions distinguishing the Kerr isolated horizon}
\centerline{\it The vacuum, Petrov type D isolated horizons}

The geometry of an isolated horizon  $(\IH, [\ell])$ can not be
assigned a `Petrov type' because it does not determine all of the
components of the Weyl tensor on $\IH$. However, if we assume that
all the components of the Ricci tensor and their first radial
derivatives vanish on $\IH$, then, in the non-extremal case, the
Bianchi identities and $(q_{ab},\D_a)$ determine the evolution of
the missing Weyl tensor component along $\IH$. Combined with the
assumption that the Weyl tensor is of the Petrov type $D$ on $\IH$
the evolution equation reduces to a certain condition on the
geometry of the isolated horizon. To write down the condition,
introduce a complex, null vector field $m$ tangent to $\IH$,
\be m^a m_a = 0,\ \ \bar{m}^a m_a\ =\ 1,
\ee
such that it is tangent to some 2 sub-surfaces in $\IH$, that is
such that
\be \Lie_{\bar{m}}m^a\ =\ (\a-\bar{\b})m^a -
(\bar{\a}-\b)\bar{m}^a. \ee
We will  denote the differential operator corresponding to the
vector field $m^a$ by $\m$,
\be
\m := m^a \partial_a.
\ee
The condition on the geometry of the isolated horizon reads
\be\label{c1a} 3\Psi_2 \mb\mb\Psi_2 + 3(\alpha
-\bar{\beta})\Psi_2\mb\Psi_2 -4 (\mb\Psi_2)^2\ =\ 0. \ee
If $\Psi_2$ vanishes at a point of $\IH$, then Weyl tensor is of
the Petrov type III, N or 0. Therefore, we can assume that
$\Psi_2\not=0$ and conclude that:
\medskip

\noindent{\bf Lemma.} {\it Suppose $(\IH, [\ell])$ is a
non-extremal isolated horizon, and  (i) the Ricci tensor and its
first radial derivative vanish on $\IH$, and (ii) the Weyl tensor
is of the type $D$ on $\IH$; then the invariant $\Psi_2$  of the
geometry of $\IH$  satisfies the following equation,
\be\label{c1b} (\mb + \a-\bar{\b})\mb(\Psi_2^{-\frac{1}{3}})\ =\
0.
\ee}
\medskip

The converse statement requires an additional assumption that,
there is an extension of the isolated horizon vector field
$\ell^a$ to a neighborhood of $\IH$ such that, the Weyl tensor
$C^a{}_{abc}$ is Lie dragged by $\ell^a$ on $\IH$. Then, the
conditions $(i)$ and (\ref{c1b}) of Lemma imply that the Weyl
tensor is of the Petrov type D at $\IH$. The above equation
(\ref{c1b}) was derived in \cite{LP}. It is independent of the
choice of a null frame $m^a,\,\bar{m}^a,\,n^a,\,\ell^a$, provided
$\ell^a$ is tangent to $\IH$ and $\Re\, m^a, \Im\, m^a$ are
surface forming at $\IH$. Notice also that, it involves the 4th
order derivatives of the 2-metric $q_{ab}$, because \be
-2\Re\Psi_2\ =\ \m (\a-\bar{\b}) + \mb (\bar{\a}-\b) -
2(\a-\bar{\b})\overline{(\a-\bar{\b})}. \ee (The condition $(i)$
can be weakened \cite{LP}:  not all of the Ricci tensor components
have to satisfy $(i)$.)
\medskip

\noindent{\bf Definition} {\it  A non-extremal isolated horizon is
vacuum, type D, whenever its geometry  satisfies the condition
(\ref{c1b}).}
\medskip

\centerline{\it The conditions distinguishing the Kerr isolated horizon}

We say that an isolated horizon $(\IH, [\ell])$ admits an axial
symmetry whenever it admits a  vector field $\Phi^a$ tangent to
$\IH$,  whose all the orbits are closed, and such
that\footnote{For $\Phi$ to be a symmetry of the geometry, it is
enough if the second equation is weakened to $\Lie_\Phi\ell^a =
a_0 \ell^a$, $a_0$ being a constant. But in the case of an axial
symmetry, $a_0=0$ necessarily.} \be \L_\Phi q_{ab}\ =\ 0, \ \
\Lie_\Phi \ell^a\ =\ 0, \ \  [\Lie_\Phi, \D_a]=0. \ee

\medskip

\noindent{\bf Proposition} {\it Suppose  a non-extremal isolated
horizon $(\IH, [\ell], q_{ab},\D_a)$ admits an axial symmetry
group generated by a vector field $\Phi^a$; then it is vacuum,
type D if and only if the following equation is satisfied,
\be\label{c2} d(\Psi_2^{-\frac{1}{3}})\ =\ A_0\Phi\vins \vol, \ee
where $d$ is the exterior derivative on $\IH$ and $A_0$ is a
complex constant which turns out to be pure imaginary.}
\medskip

The condition (\ref{c1b}) (as well as (\ref{c2})) is  a complex
equation on the metric tensor $q_{ab}$ and the rotation scalar
$\Omega$. We  found all  local solutions $(q_{ab}, \Omega)$
defined in some open subset of $\IH$. Imposing on $\Psi_2$ the
globality condition (\ref{int}) restricts the set of solutions to
a two dimensional family $(q_{ab}^{(A,J)},\Omega^{(A,J)})$,
parametrized by two real parameters $A > 0, J\ge 0$. The
parameters have a geometrical and physical  meaning, namely their
values are equal to the  area  and, respectively, the angular
momentum \cite{ABL3} of the corresponding non-extremal isolated
horizon $(\IH^{(A,J)}, [\ell^{(A,J)}])$. Given an axi-symmetric,
non-extremal vacuum, type D isolated horizon $(\IH, [\ell])$, to
find the corresponding values of $(A,J)$, one has to use the
following quasi-local formulas involving an arbitrary
cross-section $\tIH$ of $\IH$,
\ba A &=&  A_\IH = \int_{\tIH} \vol\label{A}\\
J &=& |J_\IH|  =  \frac{1}{4\pi} |\int_{\tIH} \phi\, \Im\, \Psi_2\,
\vol\label{J}|\\
\ea
where the function $\phi$  is defined up an additive constant as
the generator of the vector field $\Phi$, that is
\be  \phi_{,a}\ :=\ \Phi^a\,\vol_{ab}\ee
Now, $(\IH^{(A,J)}, [\ell^{(A,J)}])$ is the  Kerr outer isolated
horizon provided
\be \frac{A}{8\pi}\ >\ J,
 \ee
and  the Kerr inner horizon if
\be \frac{A}{8\pi}\ <\ J. \ee
However, in the case when
\be \frac{A}{8\pi}\ =\ J \ee
the pair $(q_{ab}^{(A,J)},\Omega^{(A,J)})$ coincides with the
metric tensor and the rotation scalar induced on the event horizon
of the Kerr solution in  the extremal case. On the other hand, the
corresponding  $(q_{ab}^{(A,8\pi A)}, \Omega^{(A, 8\pi A)})$ is
non-extremal by definition; let as call it a {\it special isolated
horizon}.\footnote{The geometry of every special isolated horizon
$\IH$ has a certain non-generic property \cite{ABL2,LP2}: it
admits a 2-dimensional family of null symmetries; each generator
defines a distinct isolated horizon structure on $\IH$ and exactly
one of them is extremal.}   We can conclude our results by the
following Theorem.
\medskip

\noindent{\bf Theorem} {\it Suppose  $(\IH, [\ell])$ is a vacuum,
axi-symmetric, non-extremal isolated horizon such that $A_\IH >
8\pi J_\IH$ \ (respectively, $A_\IH < 8\pi J_\IH$), where $A_\IH$
is the area and $J_\IH$ is the angular monentum of $\IH$. Then
each of the following two properties implies that $(\IH, [\ell])$
is the Kerr outer (inner) isolated horizon:
\begin{itemize}
\item[(i)] it is vacuum, type D; i.e is it satisfies (\ref{c1b})
\item[(ii)] it satisfies the condition (\ref{c2})
\end{itemize}
Conversely, every non-extremal Kerr outer (inner) isolated horizon
satisfies all the properties assumed above.}
\medskip

 \noindent{\bf Remark 1.} A priori
it may happen, that the geometry $(q_{ab},\D_a)$ of an isolated
horizon $(\IH, [\ell])$ admits an axial symmetry which does not
commute with the null flow of $[\ell]$. Then, we can pick  any
$\ell^a\in [\ell]$ and average it with respect to the symmetry
group. It can be shown by using the results of \cite{ABL2}, that
the result $\bar{\ell}$ is nontrivial on every null generator of
$\IH$. If $\ell$ in non-extremal, then so is $\bar{\ell}$. More
over, the condition, that $(\IH, [\ell])$ be vacuum, type D is
expressed by the invariants independent of $[\ell]$. Hence,  if
$\IH$ admits two distinct non-extremal isolated horizon structures
$[\ell]$ and $[\ell']$, then one of them is vacuum, type D if and
only if so is the other one. In conclusion, if we drop the assumption
$\Lie_\Phi\ell^a=0$ from the definition of the axi-symmetry, then
the  Theorem still holds.
\medskip

\noindent{\bf Remark 2.} The assumption that the Weyl tensor is of
the Petrov type D excludes the vanishing of $\Psi_2$ at any point.
One could weaken this condition, and allow for the vanishing of
$\Psi_2$. However, there are no such solutions defined globally on
$\IH$.
\medskip

\noindent{\bf Remark 3.} The condition $A_\IH > 8\pi J_\IH$ in
Theorem can be replaced by the following inequality to be
satisfied at a point $x_0$ belonging to the symmetry axis,
 \be\label{ine}
   \left| \Re\big[(\Psi_2)^{-\frac{1}{3}}(x_0)\big] \right|
 > \left| \Im \big[(\Psi_2)^{-\frac{1}{3}}(x_0)\big] \right|, \ee
and the conclusion still holds.

\section{Applications of the result}\label{A}
The results of the previous section are relevant for the
comparison of the space-time metric tensor near a non-expanding
horizon $\IH$ with the Kerr metric. For that we need certain
generalization of the Bondi coordinates \cite{ABL3}.

Every null vector field $\ell^a$ tangent to $\IH$ and every
foliation of $\IH$ with space-like 2-surfaces preserved by the
flow  $[\ell]$, defines uniquely another null vector field $n^a$
orthogonal to the leaves of the foliation and such that
$\ell^an_a=-1$.  Use the flow of $n^a$ to extend the vector field
$\ell^a$ to  a neighborhood of $\IH$. It is not any longer null
but for the simplicity let as denote it by the same letter
$\ell^a$. Then, it is true \cite{Racz} that
\be
\Lie_\ell g_{ab}{}_{\mid_\IH}\ =\ 0.
\ee
Suppose that the vacuum Einstein equations hold in a neighborhood of a
non-expanding horizon $\IH$ and that $\ell^a$ is future pointing, it
does not vanish in the future, and $\kl = \const >0$.  Then, every
transversal derivative $(\Lie_n)^m\,\, g_{ab}{}_{\mid_\IH}$,
$m=1,2,...,k,...$ exponentially converges in future to some value,
$(\Lie_n)^m\,\, g^{(\infty)}_{ab}$ say, as we move along each generator of
$\IH$ \cite{ABL3}. The values $(\Lie_n)^m\,\, g^{(\infty)}_{ab}$ are
determined by the the geometry $(q_{ab}, \D_a)$ of $\IH$ for every
$m$.  Suppose now, that $(q_{ab}, \D_a, [\ell])$ is axi-symmetric and
that the conditions
\ba \label{inv}(\mb +
\a-\bar{\b})\mb(\Psi_2^{-\frac{1}{3}})_{\mid_\IH}\ &=&\
0,\\
A_\IH\ - \ 8\pi J_\IH\ &>&\ 0 \ea
hold on $\IH$ (the inequality can be replaced by (\ref{ine})).
Then, it follows from Theorem that, all the asymptotic values
$(\Lie_n)^m\,\, g^{(\infty)}_{ab}$ of the transversal derivatives of
the metric coincide with the corresponding derivatives of the Kerr
solution.
%
Notice finally, that the quantity

\be {\cal I}\ :=\ |(\mb + \a-\bar{\b})\mb(\Psi_2^{-\frac{1}{3}})|
\ee is independent of the choice of a null frame at $\IH$,
provided $\ell$ is tangent to the null generators of $\IH$.
Therefore, it is an invariant of the geometry $(q_{ab},\D_a)$ of a
non-expanding horizon. If ${\cal I}$ fails to be zero, then its
value is a measure of the departure of the values of
$(\Lie_n)^mg^{(\infty)}_{ab}$, $m=0,1,2,...,k...$ from those of
the Kerr solution at the future, outer event horizon. These
results should be useful to numerical relativity in analyzing the
sense in which the final black hole horizon produced by a collapse
or a merger approaches the Kerr horizon.

These results should be directly applicable to numerical simulations
of black hole collisions to verify whether or not a Kerr horizon is
produced at late times and, if it is not, to estimate how large the
departure is.  The criterion involves fields defined just on the 
`world-tube' of apparent horizons and some of the existing codes  
can easily calculate them.  

\medskip

\noindent{\bf Acknowledgements} The idea of this paper  came from
Abhay Ashtekar, who also pointed out to as that, the area and the
angular momentum  of an isolated horizon can be used to
distinguish the Kerr outer and inner isolated horizons from the
special one obtained above. We have  benefited a lot from the
discussions with Jacek Jezierski, Chris Beetle, Mariusz Mroczek,
and Jerzy Kijowski. This research was supported in part by Albert
Einstein MPI, NSF grants PHY-0090091, PHY-97-34871 and the Polish
Committee for Scientific Research under grant no. 2 P03B 060 17.

\end{document}